\documentclass[prl,aps,twocolumn,twoside,superscriptaddress]{revtex4}

\usepackage{amssymb}
\usepackage{amsmath}
\usepackage{amscd}
\usepackage{latexsym}
\usepackage{epsfig}
\usepackage{graphicx}
\usepackage{bbm}

\newcommand{\qed}{{\hfill$\Box$}}

\def\bi{\begin{itemize}}
\def\ei{\end{itemize}}
\def\be{\begin{equation}}
\def\ee{\end{equation}}
\def\bea{\begin{eqnarray}}
\def\eea{\end{eqnarray}}
\def\ben{\begin{eqnarray*}}
\def\een{\end{eqnarray*}}

\def\>{\rangle}
\def\<{\langle}

\def\bo{{\bf 0}}

\def\bbZ{\mathbb{Z}}

\newcommand{\ket}[1]{| #1 \rangle}

\def\*{\star}

\def\tilde{\widetilde}

		 \def\cE{{\cal E}}

		 \def\cQ{{\cal Q}}		 

\def\cS{{\cal S}}

\begin{document}

\title{Efficiently implementable codes for quantum key expansion}
\date{\today}
\author{Zhicheng Luo}
\email{zluo@usc.edu}
\affiliation{Physics Department, University of Southern California, Los Angeles, CA 90089, USA}
\author{Igor Devetak}
\email{devetak@usc.edu}
\affiliation{Electrical Engineering Department, University of Southern California, Los Angeles, CA 90089, USA}


\begin{abstract}
The Shor-Preskill proof of the security of the BB84 quantum key distribution protocol
relies on the theoretical existence of good classical error-correcting codes with the ``dual-containing'' 
property. A practical implementation of BB84 thus requires explicit and efficiently decodable
constructions of such codes, which are not known. On the other hand, modern coding theory 
abounds with non-dual-containing codes with excellent performance and efficient decoding algorithms.
We show that the dual-containing constraint can be lifted 
at a small price: instead of a key distribution protocol, an efficiently implementable key \emph{expansion} protocol 
is obtained, capable of increasing the size of a pre-shared key by a constant factor.


\end{abstract}

\maketitle

Quantum key distribution (QKD) allows two distant parties Alice and Bob to
establish a secret key using one-way quantum communication and public classical
communication. This key is provably secure from an all-powerful eavesdropper Eve,
who is allowed to intercept the quantum communication, perform block processing of quantum 
data, and listen to the public discussion.
In contrast, key distribution by public communication alone is impossible.
QKD owes its security to two facts: 1) Alice and Bob, by performing
tomography on their (quantum) data, automatically obtain information about
Eve's (quantum) data; 2) with this knowledge Alice and Bob can perform 
information reconciliation (IR) and privacy amplification (PA) to distill
a key which is common to both (by IR), and about which Eve knows next to nothing
(by PA). In this Letter we solve the practical question of
constructing efficiently implementable codes for IR and PA.


The best known QKD protocol, BB84, was proposed by Bennett and 
Brassard in \cite{BB84}. BB84
is a simple ``prepare-and-measure'' protocol 
which can be implemented without a quantum computer or quantum memory.
Alice encodes a random bit either in the $Z$ basis 
$\{ \ket{0},\ket{1} \}$ or $X$ basis $\{\ket{+}, \ket{-} \}$
[here $\ket{\pm} = \frac{1}{\sqrt{2}} (\ket{0} \pm \ket{1})$] of a qubit system,
and sends it to
Bob. Bob performs a measurement in one of the two bases, chosen
at random. After repeating this many times, they determine by public
discussion which bits they chose the same basis for, thus establishing
a raw key. They perform channel estimation on a small fraction of the
raw key bits. If the channel is too noisy, they abort the protocol.
Otherwise they perform IR and PA on the remaining raw key bits to obtain
the final secret key.

Shor and Preskill~\cite{ShorPreskill} gave the first simple proof of the security of 
standard BB84, by relating the IR and PA steps to Calderbank-Shor-Steane (CSS) 
quantum error correcting codes. 
A CSS code protects $m$ qubits from errors by ``rotating'' them into a $2^m$
dimensional subspace of an $n$ qubit system. This subspace is the simultaneous
eigenspace of ``stabilizer'' operators of the form
\begin{eqnarray*}
Z^h & =& Z^{h(1)} \otimes \dots \otimes  Z^{h(n)}, \\
X^g & =& X^{g(1)} \otimes \dots \otimes X^{g(n)}.
\end{eqnarray*}
Here $Z$ and $X$ are Pauli matrices, and
$h = h(1) \dots h(n)$ and $g = g(1) \dots g(n)$ are  binary vectors of length $n$.
The vectors $h$ are $g$ are chosen to be rows of the classical ``parity check'' matrices
$H_1$ and $H_2$, respectively. To ensure that the stabilizer operators commute,
$H_1$ and $H_2$ must be mutually orthogonal: $H_1 H_2^T = 0$. This condition is equivalent
to saying that the codes corresponding to $H_1$ and $H_2$ contain each other's duals.
 Let the $(n-k_i) \times n$
parity check matrix $H_i$ correspond to an $[n,k_i, d]$ classical error correcting code 
which encodes $k_i$ bits into $n$ bits and corrects errors on any $t = (d-1)/2$ bits.
Then $m = k_1 + k_2 - n$ and the CSS code corrects quantum errors on any $t$ qubits.

In order to securely implement the BB84 protocol we need to find good mutually dual containing codes
of large blocklength $n$. These are known
to exist in principle, by the Gilbert-Varshamov bound for CSS codes \cite{CSS}.
Unfortunately, no explicit constructions are known, let alone ones that would be simple to decode.
Our main result is that the dual-containing  condition may be lifted.
This permits us to employ excellent efficiently decodable modern classical
codes such as LDPC \cite{G63} and turbo codes \cite{BGT93}. The price we have to pay is that
our protocol performs expansion of a pre-shared key rather than creating one
from scratch. This is not much of a drawback, as existing QKD protocols
require a logarithmic amount of pre-shared key to authenticate the
public discussion. Still we choose to make this distinction, as in our case the pre-shared
key is linear in the quantum communication cost.
 Our construction is closely related to the entanglement-assisted quantum 
codes of Brun, Devetak and Hsieh \cite{cataqecc}, which 
generalize stabilizer codes to the communication setting where the sender and receiver
have access to pre-shared entanglement.

First we consider an idealized setting in which  the eavesdropper
is known to have introduced errors on no more than a fixed fraction of the qubits.
In other words, the channel estimation is assumed to have been successfully performed.
We show how to construct an $[n,m-c,d;c]$ quantum key expansion (QKE) protocol, which expands 
the key from $c$ to $m$ bits if at most  $t = (d-1)/2$ out of $n$ qubits have become corrupted.
Then we invoke standard results \cite{LoChau,ShorPreskill,GL03} to
incorporate the channel estimation phase.

Let $H_i$ ($i = 1,2$) be the parity check matrix
for a classical $[n,k_i,d]$ code $C_i \subset \bbZ_2^n$,
so that the rows of $H_i$ form a basis for
$C_i^\perp$. 
Consider the $(n-k_1) \times (n-k_2)$  matrix $M = H_1 H_2^T$. 
In general $M \neq 0$ 
and it can be row and column reduced (i.e.
multiplied from the left and right by non-singular transformation matrices
$T_1$ and $T_2^T$) to a matrix of the form
$$
T_1MT_2^T = \left(
   \begin{aligned}
   & 0_{\ell_1\times \ell_2}  &0_{\ell_1 \times c} \\
   & 0_{ c \times \ell_2}  & \!\!I_{c\phantom{\times c}} 
   \end{aligned}\right) = J_1 J_2^T,\,\,
$$
where $n - k_1 =  c + \ell_1$ and $n - k_2 = c + \ell_2$ and
$$
J_i = \left(
   \begin{aligned}
   & 0_{\ell_i\times c}  \\
   & \,\, I_{c\phantom{\times c}}  
   \end{aligned}\right).
$$
This is the well known Gaussian elimination procedure.
Letting $\tilde{H}_i = T_i H_i$ be an equivalent parity matrix for the code $C_i$, 
we have $\tilde{H}_1 \tilde{H}_2^T = J_1 J_2^T$. 
Hence $H'_1 H^{'T}_2 = 0$, where 
 the  $(n - k_i)\times(n + c)$ ``augmented'' parity check matrices $H'_i$ (cf. \cite{cataqecc})  
are given  by
$
H'_i = ( \tilde{H}_i \,\, J_i ).
$
Note that $H'_i$  can be viewed as the parity check matrix of a classical code
$C'_i$  by defining the row space of $H'_i$ to 
be $C^{'\perp}_i$.
Then $C_2^{'\bot}\subseteq C'_1$ and $C_1^{'\bot}\subseteq C'_2$. 
 The set of errors $H'_i$  can correct is the same as the set of errors $H_i$
 can correct, assuming that the last $c$ bits are error free.
Thus $H'_i$ can correct the error set $\cS(n,c,d)$ defined 
as the set of binary row vectors of dimension $n + c$ with 
$\leq (d-1)/2$ ones among the first $n$ bits, and zeros elsewhere.
 

\begin{figure}
  \includegraphics[width=8cm]{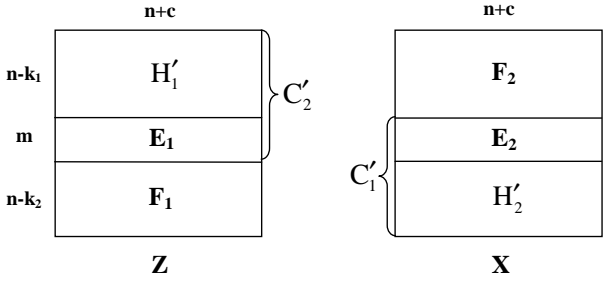}
  \caption{The construction of the full rank matrices $N_1$ and $N_2$.} 
\end{figure} 

To relate the error correcting properties of these classical codes to those of the 
corresponding CSS codes, we find it useful to extend
the parity check matrices $H'_1$ and $H'_2$ to full rank matrices
as  shown in Figure 1. Starting with $H'_1$, whose rows are a basis for $C_1^{'\perp}$,
add  $m = k_1 + k_2 + c - n$ independent row vectors 
(comprising the matrix $E_1$)
such that the rows of $H'_1$ and $E_1$ together form a basis for $C'_2 \supseteq C_1^{'\perp}$.
Collect the remaining $n - k_2$ independent vectors in the matrix $F_1$.
The $(n + c) \times (n + c)$ matrix 
$$
N_1 =  \left(
   \begin{aligned}
   & H'_1 \\
   & E_1 \\
    & F_1  
   \end{aligned}\right)
$$
has full rank, and hence so does $N_1 N_1^T$. By Gaussian elimination,
there exists a matrix $T$ such that $ N_1 N_1^T T^T = I$.
Decompose $N_2 = T N_1$  into three segments just like $N_1$:
$$
N_2 =  \left(
   \begin{aligned}
   & F_2  \\
   & E_2 \\
    & \hat{H}_2  
   \end{aligned}\right).
$$
The condition $N_1 N_2^T = I$ is now written as 
\bea
\label{relations}
H'_1 F_2^T = I, \quad  H'_1 E_2^T = 0, \quad  H'_1 \hat{H}_2^T = 0, \nonumber \\
E_1 F_2^T = 0, \quad  E_1 E_2^T = I, \quad  E_1 \hat{H}_2^T = 0,  \\
F_1 F_2^T = 0, \quad  F_1 E_2^T = 0, \quad  F_1  \hat{H}_2^T = I.\nonumber 
\eea
Hence the rows of $\hat{H}_2$ form a basis for $C^{' \perp}_2$.
With an appropriate redefinition of $F_1$ we can identify  
$\hat{H}_2$ with  $H'_2$. $H'_2$ and $E_2$ together form a basis for 
$C'_1 \supseteq C_2^{'\perp}$.

An error set $\cE_1 \subset \bbZ_2^{n +c}$ 
correctable by the code $H'_1$ is of the form
$$
\cE_1 = \{ b F_2 + \beta(b) E_2 + \beta'(b) H'_2: 
b \in \bbZ_2^{n - k_1} \},
$$ where
$\beta: \bbZ_2^{n - k_1} \rightarrow \bbZ_2^{m}$
and $\beta':  \bbZ_2^{n - k_1} \rightarrow \bbZ_2^{n - k_2}$
are known functions. Thus $\beta(b)$ is 
a row vector of dimension $m$ and $\beta(b) E_2$ is 
an element of the row space of $E_2$.
$b$ is called the \emph{error syndrome} since it uniquely specifies the error.
For an error $u \in \cE_1$,  by (\ref{relations}), the error syndrome is calculated as $b = H'_1 u^T$.  
Since $H_1$ is an $[n,k_1,d]$ code,  $\cS(n,c,d) \subseteq \cE_1$. 

Similarly, an error set $\cE_2 \subset \bbZ_2^{n +c}$ 
correctable by the code $H'_2$ is of the form
$$
\cE_2 = \{ p F_1 + \alpha(p) E_1 + \alpha'(p) H'_1: 
p \in \bbZ_2^{n - k_2} \},
$$ where
$\alpha: \bbZ_2^{n - k_2} \rightarrow \bbZ_2^{m}$
and $\alpha':  \bbZ_2^{n - k_2} \rightarrow \bbZ_2^{n - k_1}$
are known functions.
Since $H_2$ is an $[n,k_2,d]$ code, $\cS(n,c,d) \subseteq \cE_2$.


In the Shor-Preskill proof \cite{ShorPreskill} a QKD protocol was obtained
by modifying an entanglement distillation protocol.
Our starting point is an \emph{entanglement assisted} entanglement distillation (EAED) protocol. 
Alice and Bob initially share the state  $\ket{\Phi}^{\otimes c}$, where
$$
\ket{\Phi}^{AB} = \frac{1}{\sqrt 2} (\ket{0}^A \ket{0}^B + \ket{1}^A \ket{1}^B)
$$
is the \emph{ebit} state.

Their goal is to distill a total of $m$ ebits. The resources at their disposal are
classical communication and a noisy $n$-qubit channel, 
which introduces errors on at most $t$ out of $n$ qubits.
At the beginning of the protocol, Alice creates another $n$ ebit states 
locally and sends the $B$ part of them through the noisy channel to Bob.
An operator written as $U \otimes V$ means that $U$ acts 
on ``subsystem $A$'', the $n + c$ qubits that stay with Alice,
 and $V$ acts on ``subsystem $B$'', the $n + c$ qubits
which end up in Bob's possession.
We will describe the noise more generally as acting on the latter $n+c$ qubits 
(even though only the first $n$ of these are affected):
let $\cQ(n,c,d)$ be the set of 
error operators of the form $I \otimes (X^{h_1} Z^{h_2})$
where $h_1, h_2 \in \cS(n,c,d)$. 
The $X$-type errors are called \emph{bit errors} and the $Z$-type
errors are called \emph{phase errors}. 
Because $\cS(n,c,d) \subseteq \cE_1\cap \cE_2$,
every element of  $\cQ(n,c,d)$ is of the form 
\be
 I \otimes 
(Z^{p F_1} X^{b F_2} Z^{\alpha(p) E_1} Z^{\alpha'(p)H'_1}  
X^{\beta(b) E_2} X^{\beta'(b) H'_2} 
),
\label{form}
\ee
for some 
$p \in \bbZ_2^{n - k_1}, b \in \bbZ_2^{n - k_2}$. 


\vspace{1mm}

Our EAED protocol comprises the following steps:
\begin{enumerate}
	\item The initial state is $\ket{\Phi}^{\otimes n + c}$.
The first $n$ ebits are held entirely
by Alice, and the last $c$ are shared between Alice and Bob.
Denote by $\bo$ the [$(n + c)$-dimensional] all-zero vector. 
The state $\ket{\Phi}^{\otimes n + c}$ is the simultaneous $(-1)^{(\bo,\bo)}$ eigenstate of 
$\{Z^{I_{n+c}} \otimes Z^{I_{n+c}}, X^{I_{n+c}}\otimes X^{I_{n+c}}\}$. By this we mean that 
it is the $(-1)^{0} = 1$ eigenstate of $Z^{e}\otimes Z^{e}$ for each row $e$ of 
the $(n+c) \times (n+c)$ identity matrix ${I_{n+c}}$.
As the matrices $N_1$ and $N_2$ are full rank, 
this state is equivalently described as the 
simultaneous $(-1)^{(\bo,\bo;\bo,\bo;\bo,\bo)}$ eigenstate of the operators 
\bea
\{Z^{H'_1}\otimes Z^{H'_1}, X^{F_2}\otimes X^{F_2}; \nonumber \\
 Z^{E_1}\otimes Z^{E_1}, X^{E_2}\otimes X^{E_2}; \label{cg}\\ 
Z^{F_1}\otimes Z^{F_1}, X^{H'_2}\otimes X^{H'_2}\}. \nonumber
\eea
In other words, it is the $(-1)^{0}$ eigenstate of 
$Z^{h_1}\otimes Z^{h_1}$ for each row $h_1$ of $H'_1$, etc.

\item An error in $\cQ(n,c,d)$ of the form (\ref{form})
occurs.
The new state is the simultaneous 
$(-1)^{(b, \alpha'(p); \beta(b), \alpha(p); \beta'(b), p)}$ 
eigenstate of the operators in (\ref{cg}). 	
This is easily seen from the relations (\ref{relations})
and the fact that acting with $(I \otimes X^g)$ on a eigenstate
of $(Z^h \otimes Z^h)$ with eigenvalue $(-1)^a$, changes the
eigenvalue to $(-1)^{a + gh^T}$ [and similarly with $X$ and $Z$ interchanged].
	\item  In order to find out the error syndromes $b$ and $p$, Alice
and Bob should measure the commuting operators 
$\{ Z^{H'_1}\otimes Z^{H'_1},  X^{H'_2}\otimes  X^{H'_2} \}$.
However, this would require a non-local measurement. 
Since $Z^{h} \otimes Z^{h} = (Z^{h} \otimes  I) (I \otimes Z^{h})$,
Alice and Bob can effectively measure $ Z^{h}\otimes Z^{h}$ by 
Alice measuring $Z^{h}\otimes I$, Bob measuring 
$I\otimes Z^{h}$ and multiplying the measurement outcomes.
Thus, Alice measures $Z^{H'_1}\otimes I$ and $X^{H'_2}\otimes I$, obtaining 
$b'$ and $p'$, and Bob measures 
	        $I\otimes Z^{H'_1}$ and  $I\otimes X^{H'_2}$, obtaining $b''$ and $p''$. 
Alice sends Bob her measurement outcomes and Bob computes $p = p' + p''$ and $b = b' + b''$. 
 \item Bob performs the correction operation $I \otimes ( Z^{\alpha(p) E_1} 
X^{\beta(b) E_2})$.
Alice and Bob are left with the simultaneous $(-1)^{(\bo,\bo)}$ eigenstate
of $\{ Z^{E_1} \otimes Z^{E_1}, X^{E_2} \otimes X^{E_2} \}$.
They can transform this state by local unitaries into $\ket{\Phi}^{\otimes m}$.
\end{enumerate}

This EAED protocol is readily made into an entanglement-assisted
secret key distillation protocol.
Starting with the distilled  state $\ket{\Phi}^{\otimes m}$,
Alice measures $Z^{I_m} \otimes I$ and Bob measures $I \otimes Z^{I_m}$ to 
obtain a common key $k \in \bbZ_2^{m}$. The key is decoupled from the rest of the world,
and hence Eve,
because $\ket{\Phi}^{\otimes m}$ is a pure state.

We proceed to simplify this key distillation protocol.
Instead of transforming into $\ket{\Phi}^{\otimes m}$ in step 4,
and measuring $\{ Z^{I_m} \otimes I, I \otimes Z^{I_m}\}$,
it suffices to measure $Z^{E_1} \otimes I$ and  $I \otimes Z^{E_1}$ to obtain $k$ directly.
In step 4, Bob need not perform the phase error part  
$I \otimes  Z^{\alpha(p)E_1}$ of the correction operation; this commutes with 
the $Z^{E_1} \otimes I$ and $I \otimes Z^{E_1}$ operators, and hence does
not affect the measured key value $k$. Thus measuring  $X^{H'_2}\otimes I$
and $ I\otimes X^{H'_2}$ in step 3 is also unnecessary. 
Bob performing the bit error correction $I \otimes  X^{\beta(b)E_2}$,
followed by measuring $I \otimes Z^{E_1}$ to get $k$, is equivalent to just measuring
$I \otimes Z^{E_1}$ to get $k'$ and computing $k = k' + \beta(b)$.

The new key distillation protocol consists of steps 1 and 2, followed by:

\begin{itemize} 
 \item [3.] Alice measures $Z^{H'_1}\otimes I$, obtaining 
$b'$, and Bob measures $I\otimes Z^{H'_1}$, obtaining $b''$. 
Alice sends $b'$ to Bob. 
 \item [4.] Alice measures $Z^{E_1} \otimes I$, obtaining $k$, and Bob measures
$I \otimes Z^{E_1}$, obtaining $k'$. Bob computes $k = k' + \beta(b' + b'')$.
\end{itemize}

The above protocol requires multi-qubit operations and pre-shared entanglement.
We will now reduce it to single-qubit operations and replace 
the entanglement by a pre-shared secret key.

As they commute with all the other steps, Alice may perform
her $Z^{H'_1}\otimes I$ and  $Z^{E_1}\otimes I$ measurements  before step 2.
For the same reason she can measure $Z^{F_1}\otimes I$ at the same time.
Together, these three measurements are equivalent to Alice  measuring 
the $Z$ operator of each individual qubit, obtaining a string $u \in \bbZ_2^{n + c}$. Then 
$b' = H'_1 u^T$ and $k = E_1 u^T$.

Bob can measure $I \otimes Z^{F_1}$ at the end of the protocol, as
it commutes with $I \otimes Z^{H'_1}$ and $I \otimes Z^{E_1}$.
Together, these three measurements are equivalent to
Bob  measuring the $Z$ operator of each individual qubit.
The measurement result $v  \in \bbZ_2^{n + c}$
is used to compute $b'' = H'_1 v^T$ and $k' = E_1 v^T$.

Because Bob has the last $c$ qubits from the beginning 
(equivalently, the noise acts as the identity on them), he can
measure them at the same time as Alice.  
Alice and Bob performing local $Z$ measurements on the last $c$ ebits
is as if they shared a secret key $\kappa$ of length $c$ bits to start with.
Alice measuring half of an ebit $\ket{\Phi}$ in the $Z$ basis and sending the other half 
through the channel is equivalent to her preparing $\ket{0}$ or $\ket{1}$ at random and 
sending it through the channel.
This leaves us with the $[n, k_1 + k_2 - n, d;c]$ QKE protocol below.

\begin{enumerate}
	\item Alice and Bob share a secret key $\kappa$ of length $c$ bits.
Alice generates a random $n$ bit string $r$. Together they form the
$n + c$ bit string $u = (r,\kappa)$. 
Alice computes $b' = H'_1 u^T$ and $k = E_1 u^T$.
\item Alice prepares the product state $\ket{r}$ in 
her lab and sends it over the noisy channel.
\item Bob receives the corrupted $n$ qubit state. He measures each qubit in the $Z$
basis, obtaining  a string $r'$, which together with Bob's copy of the initial secret key $\kappa$
forms the $n + c$ bit string $v = (r',\kappa)$. 
Bob computes $b'' = H'_1 v^T$ and $k' = E_1 v^T$.
\item Alice sends $b'$ to Bob. Bob computes $k = k' + \beta(b' + b'')$.
\end{enumerate}

The above  QKE protocol deals with the unrealistic situation in which
  Eve is known to have introduced no more than $t$ errors.
To deal with the most general eavesdropping attack, the so-called coherent attack,
Alice and Bob need to be able to estimate the effective channel introduced by Eve. 
It was shown in \cite{GL03} that there is no loss of generality in assuming that
Eve effects a Pauli channel, i.e. one that applies elements of the Pauli group
chosen with particular probabilities. However, preparing and measuring only
in the $Z$ basis is insufficient to estimate the channel; for instance, phase errors pass undetected.
The BB84 protocol circumvents this problem by preparing and measuring in both the $Z$ and $X$ bases.
Alice will have to send a total of $(2 + 3 \delta)n$ qubits:
the factor of $2$ comes from number of different bases used and 
a small fraction $\delta n$ is reserved for channel estimation.
The details of the protocol follow:

\medskip

{ 1.} Alice creates $(2 + 3 \delta)n$ random bits.

{ 2.} Alice chooses a random $(2+ 3\delta)n$-bit string $a$, which
 determines whether the corresponding bit of $r$ is to be prepared in the 
$Z$ (if the corresponding bit of $a$ is $0$) or $X$ basis (if the bit of $a$ is $1$). 

{ 3.} Alice sends the qubits to Bob.

{ 4.} Bob receives the $(2+ 3\delta)n$ qubits and measures each
	      in the $Z$ or $X$  basis at random.

{ 5.} Alice announces $a$.

{ 6.} Bob discards any results where he measured a different
	       basis than Alice prepared in.  With high probability, there are at
	      least $(1 + \delta)n$ bits left (if not, abort the protocol).  Alice 
        randomly chooses a set of $n$ bits to be her string $r$, and Bob's corresponding
bits comprise $r'$. The remaining $n\delta$ bits are used for channel estimation. 

{ 7.} Alice and Bob publicly announce the values of their channel estimation bits.
	      If the estimated channel introduces more than $t$ errors,
they abort the protocol.

{ 8.} Alice computes $b' = H'_1 u^T$ and $k = E_1 u^T$, where $u = (r,\kappa)$.
Alice announces $b'$.

{ 9.} Bob computes $b'' = H'_1 v^T$ and $k' = E_1 v^T$, where $v = (r',\kappa)$.
Bob's estimate of the key $k$ is $k' + \beta(b' + b'')$.
 
\medskip

Observe that if the protocol fails at any point, the pre-shared key $\kappa$ remains
uncompromised. Since the protocol was obtained from an entanglement distillation protocol,
it is also universally composable \cite{Renner}.

As in \cite{ShorPreskill, GL03}, we can go beyond fixed-distance codes, and
instead use codes which merely perform well on i.i.d. (independent, identically
distributed) channels. This is achieved  by Alice performing a random permutation
on her bits, and announcing it to Bob in step 5, thus symmetrizing the noisy 
channel induced by Eve's actions. If Alice and Bob estimate 
a rate $q$ of $X$ and $Z$ errors, it suffices for $C_1$ and $C_2$ to 
perform well on a binary symmetric channel (BSC) with error parameter slightly above $q$ \cite{GL03}.
Modern classical codes such as turbo codes \cite{BGT93} and LDPC codes \cite{G63} 
can essentially achieve the Shannon capacity $1 - H(q)$ on a BSC.  Moreover, these codes are
(suboptimally) decodable in polynomial time. 
This gives a key rate of $(m-c)/n \approx 2(1 - H(q)) - 1
= 1 - 2 H(q)$, which hits $0$ for $q = 0.11$. Thus, with 
a QKE protocol based on modern codes we can tolerate the Shor-Preskill
bound $q = 0.11$ in practice. 

At present there is a large gap between abstract security proofs of QKD,
which rely on the theoretical existence of certain codes, and experimental implementations,
which use  PA and IR codes chosen ad hoc and are thus not proven to be secure.
Our result bridges this gap: it makes accessible the example
of modern turbo and LDPC codes which are readily available, easy to encode and decode, 
yet provide a basis for unconditionally secure key distribution
protocols.
The performance of specific modern codes is 
currently under investigation.

\acknowledgments We thank Graeme Smith and Todd Brun for useful discussions. This work was
supported in part by the NSF Career grant no. 0545845.

\end{document}